\documentclass[sigconf, natbib, nonacm]{acmart}
\usepackage{graphicx} % Required for inserting images
\usepackage{stfloats} 

\title{A Case Study of Next Portfolio Prediction for Mutual Funds}
\author{Guilherme Yambanis Thomaz}
\email{guilherme.thomaz@usp.br}
\affiliation{%
  \institution{University of São Paulo}
  \state{São Paulo}
  \country{Brazil}
}

\author{Denis Deratani Mauá}
\email{ddm@usp.br}
\affiliation{%
  \institution{University of São Paulo}
  \state{São Paulo}
  \country{Brazil}
}

\date{May, 2024}

\begin{document}

\begin{abstract}
Mutual funds aim to generate returns above market averages. While predicting their future portfolio allocations can bring economic advantages, the task remains challenging and largely unexplored. 
To fill that gap, this work frames mutual fund portfolio prediction as a Next Novel Basket Recommendation (NNBR) task, focusing on predicting novel items in a fund's next portfolio. We create a comprehensive benchmark dataset using publicly available data and evaluate the performance of various recommender system models on the NNBR task.
Our findings reveal that predicting novel items in mutual fund portfolios is inherently more challenging than predicting the entire portfolio or only repeated items. While state-of-the-art NBR models are outperformed by simple heuristics when considering both novel and repeated items together, autoencoder-based approaches demonstrate superior performance in predicting only new items.
The insights gained from this study highlight the importance of considering domain-specific characteristics when applying recommender systems to mutual fund portfolio prediction. The performance gap between predicting the entire portfolio or repeated items and predicting novel items underscores the complexity of the NNBR task in this domain and the need for continued research to develop more robust and adaptable models for this critical financial application.
\end{abstract}

% <concept_id>10002951.10003317.10003347.10003350</concept_id>
% <concept_desc>Information systems~Recommender systems</concept_desc>
% <concept_id>10002951.10003317.10003359</concept_id>
% <concept_desc>Information systems~Evaluation of retrieval results</concept_desc>
% <concept_significance>300</concept_significance>
% <concept_id>10010147.10010257.10010282.10010292</concept_id>
% <concept_desc>Computing methodologies~Learning from implicit feedback</concept_desc>
% <concept_significance>100</concept_significance>

% \ccsdesc[500]{Information systems~Recommender systems}
% \ccsdesc[300]{Information systems~Evaluation of retrieval results}
% \ccsdesc[100]{Computing methodologies~Learning from implicit feedback}


\maketitle

\section{Introduction}

Mutual funds are popular investment vehicles that aim to generate returns above market averages. The industry continues to grow, indicating that investors value the professional management and investment strategies offered by these funds. However, modeling these strategies and thus being able to predict the future portfolio allocations of mutual funds remains a challenging task due to the complex dynamics of fund managers' investment strategies and the multi-item nature of portfolios. 

Recommender systems aim at assisting users in discovering relevant items from a large pool of options.
These systems employ techniques such as collaborative filtering, content-based filtering, and hybrid approaches to provide personalized recommendations based on user behavior, contextual information, and profiling \cite{adomavicius2005toward}.
Recommender systems have been applied to various domains in finance, such as in recommending specific stocks to buy or sell and recommending asset allocation strategies.
Traditional recommender systems, however, treat user interactions as independent events, thus not accounting for the temporal or sequential aspects of user behavior and preferences.

Next Basket Recommendation (NBR) is an extension of the traditional recommender system paradigm that explicitly models the sequential nature of user interactions. In NBR, the goal is to predict the next set of items a user may be interested in based on their previous purchase or interaction history. 
In our targeted domain, that means predicting the set of stocks in which a mutual fund invests in the next term, given historical data on mutual fund allocation strategies.
By capturing the temporal dependencies between user actions, NBR aims to provide more context-aware and timely recommendations, leading to improved user satisfaction and business outcomes.
Novel Next Basket Recommendation (NNBR) is a recently proposed variant of NBR that focuses on recommending a set of items that have not been acquired recently by the user \cite{btbr}. 
This is particularly relevant in domains where the discovery of novel items is nontrivial or more important.

Although NBR and NNBR models have been applied to various domains, such as e-commerce and retail, their application to the financial domain, particularly to the prediction of mutual fund portfolios, remains largely unexplored.

\begin{figure}
    \includegraphics[width=\linewidth]{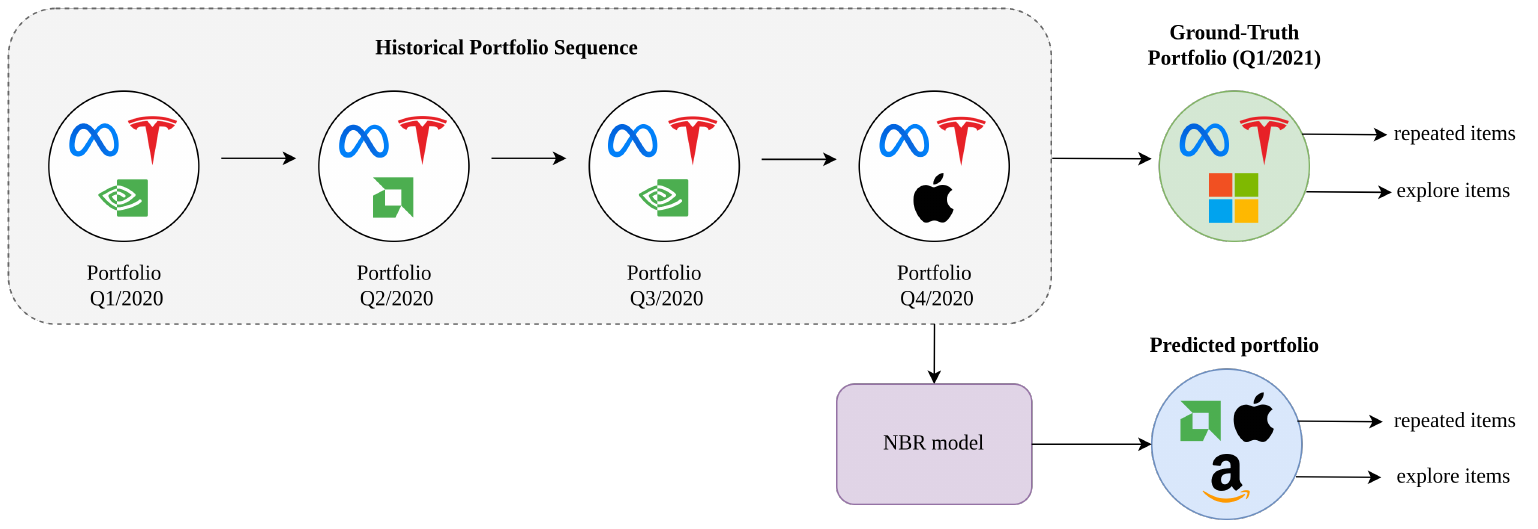}
    \Description[Schematic view of the Mutual Fund Next Portfolio Prediction Task.]{Schematic view of the Mutual Fund Next Portfolio Prediction Task.}
    \caption{Schematic view of the Mutual Fund Next Portfolio Prediction Task.}
      \label{fig:nbr_funds}
\end{figure}

This paper describes new task as well as an application of recommender systems to predict mutual fund portfolio allocations (see Figure~\ref{fig:nbr_funds}). 
In particular, we explore the use of state-of-the-art techniques for the Next Novel Basket Recommendation (NNBR) and Next Basket Repurchase Recommendation (NBRR) tasks.
By leveraging the insights gained from the application of NNBR and NBRR techniques, fund managers can enhance their investment decision-making processes, identify new opportunities, and stay ahead of market trends. 
Investors can use the techniques to compare mutual funds based on their predicted allocations, aligning their investment choices with their financial goals and risk preferences.

Unlike typical NBR domains, such as retail, where user behavior is usually more dynamic and prone to exploration, mutual fund portfolios tend to exhibit a high degree of stability over time. 
Fund managers often maintain a core set of holdings, making only incremental changes to their portfolios based on market conditions and investment objectives. This characteristic makes the application of the NNBR and NBRR techniques crucial to distinguish between repeated and novel items to uncover meaningful patterns and insights.

By focusing on predicting novel items in a fund's next portfolio, we can help identify shifts in investment strategies, potential new opportunities, or emerging trends that may not be apparent from the fund's historical holdings alone. 
The ability to distinguish between repeated and novel items is especially valuable in the mutual fund domain, where the discovery of new investment patterns can have significant implications for fund performance and risk management.

Our main contributions are:
\begin{itemize}
\item We introduce a novel application of NNBR techniques to the domain of mutual fund portfolio prediction, addressing the unique challenges and opportunities presented by this context.

\item We create a comprehensive benchmark dataset using publicly available data from the SEC, Open FIGI, and Yahoo Finance, focusing on mutual funds with at least a partial allocation in U.S.-listed stocks from 2020 to 2021.

\item We evaluate the performance of various recommender system models, including NBR models, traditional recommender systems, and autoencoders, on the NNBR task in the context of mutual fund portfolio prediction, using the RecBole library \cite{recbole} to ensure reproducibility.
\end{itemize}

Our findings reveal that state-of-the-art NBR models are outperformed by simple heuristics for predicting  mutual fund allocation, when considering both novel and repeated items together.

For the subtask of predicting only new items in the portfolio, autoencoder-based approaches outcompete all the other models in their out-of-the-box implementation. Those results highlight the importance of considering the specific characteristics of the mutual fund domain when applying recommender systems and the potential of autoencoders for predicting novel portfolio allocations.
In addition, our results show that the task of predicting novel items in mutual fund portfolios is inherently more challenging than that of predicting repeated items.

\section{Related Work}

In the context of individual stock prediction, traditional techniques, such as fundamental and technical analysis, have been widely used to support investment decisions. These methods focus on analyzing price volume patterns and financial ratios of individual stocks to identify undervalued or promising investment opportunities. This traditional methodology has also been extended to machine learning techniques. However, these approaches focus primarily on individual stocks and do not consider the relationships between stocks or the composition of a portfolio.

To address this limitation, \citeauthor{rule_minig} \cite{rule_minig} proposed a stock market recommender system based on association rule mining (ARM) that recommends a portfolio of stocks. By dividing the stock market into thematic sectors and generating cross-sector and intra-sector rules, their system aims to find correlations between stocks and recommend a diversified portfolio. This approach differs from traditional techniques by considering the interrelationships between stocks and providing portfolio-level recommendations.

In the domain of asset allocation strategy recommendation, knowl\-edge-based recommender systems (KBRS) have been identified as a promising approach due to the limitations of content-based and collaborative filtering methods in this context. \citeauthor{case-base_rec} \cite{case-base_rec} proposed a framework for recommending asset allocation strategies using case-based reasoning combined with a diversification strategy. Their framework aims to support financial advisors in proposing diverse and personalized investment portfolios.

Recommender systems have also been proposed to create personalized investment recommendations, such as in the works by \cite{IRJET2022}, \cite{XAIfundsGraphs}, and \cite{McCreadie2022}. Some works have also tried to express mutual fund similarity, such as \cite{satone2021fund2vec}. Although inspired by these works, our proposal is different, since we explicitly model mutual funds' behavior, predicting their next portfolio with a focus on novel items. By identifying the best recommender system models for the NNBR task in the mutual fund domain, we aim to provide insights that could be useful for applications such as finding funds with similar behavior or even for investors who want to replicate a fund's strategy.

While these works have made significant contributions to the application of recommender systems in finance, they have not specifically addressed the problem of predicting mutual fund portfolio allocations using Next Novel Basket Recommendation (NNBR) techniques. Our work aims to fill this gap by introducing a novel application of NNBR to the mutual fund domain, creating a comprehensive benchmark dataset, and evaluating the performance of various recommender system models on this task. By focusing on predicting novel items in a fund's next portfolio, we aim to uncover shifts in investment strategies and identify potential new opportunities that may not be apparent from historical holdings alone.

It should be noted that while the primary focus of our work is to predict the next portfolio, the intermediate outputs of the recommender system models, such as fund and stock embeddings, could also be utilized in other downstream tasks. These embeddings can capture latent relationships and similarities between funds and stocks, which may be valuable for applications such as fund clustering, portfolio diversification, or risk assessment.

In summary, Next Novel Basket Recommendation (NNBR) offers a promising approach to model the investment strategies of mutual fund managers by capturing the temporal dependencies and patterns in their portfolio holdings, with a specific focus on predicting novel items. By extending the NNBR framework to the financial domain and evaluating various recommender system models, we aim to identify the best-performing techniques for this task. The insights gained from this research can provide valuable information about the future novel holdings of mutual funds, which can have significant implications for fund managers, investors, and the broader finance industry.

To achieve this goal, we will be testing a range of recommender system models, including general recommendation models, sequential recommendation models, and Next Basket Recommendation models. Additionally, we will establish baseline performance using simple yet effective heuristics. In the following section, we will discuss these models in detail and explain how they will be applied to the NNBR task in the mutual fund domain.

\section{A New Benchmark For Mutual Fund Portfolio  Prediction}

The application of NBR techniques, specially the NNBR task, to prediction mutual fund portfolio is to our knowledge new, and thus, lacks a benchmark for evaluation of different approaches.
We now describe the process of data acquisition and processing for obtaining such a benchmark for the task at hand, which is a contribution of this work.
%To effectively apply recommender system techniques to the Next Novel Basket Recommendation (NNBR) task, to mutual fund portfolio prediction, we require a comprehensive and reliable dataset that captures the historical holdings of mutual funds. In the following section, we describe the process of creating a novel dataset specifically designed for this task, which 
%As we will describe, building such a dataset involved collecting, processing and fusing data from various financial sources such as the SEC's NPORT-P filings and the Open FIGI API, and is per se a contribution of this work.
%This dataset will serve as the foundation for our experiments and analysis, enabling us to evaluate the performance of different recommender system models in predicting mutual fund portfolios.

To create our dataset, we collected historical data from the SEC's NPORT-P filings, which are mandatory quarterly reports containing detailed information on a fund's portfolio holdings. These filings, introduced in 2018 as part of the SEC's modernization of investment company reporting, provide a comprehensive snapshot of a fund's investments, including the name, identifier, balance, and value of each holding. By requiring funds to disclose their complete portfolio holdings on a quarterly basis, NPORT-P filings aim to increase transparency and help investors make informed decisions. Although these filings offer valuable insight into fund strategies and risk exposures, it is important to note that they may be subject to certain limitations, such as the 60-day grace period for reporting and the potential for window dressing, where funds may adjust their holdings just before the reporting date to present a more favorable image.

The data extraction process involves obtaining historical N-PORTP forms through the SEC EDGAR API, which returns XML files containing data for each fund and period. These files are processed and transformed into a dataframe. To gather additional information about the assets, such as their listing on US stock exchanges and asset type, we query the Open FIGI API. This step helps to ensure the accuracy and completeness of the dataset.

We then apply a series of filters to refine the dataset. Only assets with a valid CUSIP identifier, listed on US stock markets and labeled as "Common Stocks" by Open FIGI are retained. Furthermore, we keep assets with a "Long" pay-off profile and an allocation of at least 10,000 USD. %Figure~\ref{fig:data_processing} summarizes the data transformation process.
After filtering the funds and common stock holdings, we download the corresponding time-series data for each common stock using the Yahoo Finance Python package. Incorporating these data into the modeling stage remains an area for future exploration.

% \begin{figure}
%     \includegraphics[width=\linewidth]{figuras/data_processing.pdf}
%     \caption{Summary view of data processing steps}
%       \label{fig:data_processing}
% \end{figure}

Our dataset focuses on the US stock market and specifically on common stocks. An interesting potential extension would be to consider incorporating shorted positions as negative edges and investigate how models could effectively incorporate this information.

The resulting dataset, consisting of 5,014 funds and 4,917 stocks, provides a unique opportunity to apply recommender system techniques, particularly Next Novel Basket Recommendation (NNBR), in the context of finance.

\begin{table}
\centering
\caption{Portfolio Summary Statistics for Q1-2021}
\begin{tabular}{rrrrrrrr}
\toprule
{} & {mean} & {std} & {min} & {25\%} & {50\%} & {75\%} & {max} \\
\midrule
{Portfolio Size} & 175.4 & 280.1 & 1 & 37 & 69 & 181 & 2878 \\
{\% Mean Alloc.}  & 2.5 & 6.03 & 0.03 & 0.55 & 1.45 & 2.7 & 100 \\ 
\bottomrule
\end{tabular}
\label{tab:basic_stats}
\end{table}

Table \ref{tab:basic_stats} presents the summary statistics for the portfolios in Q1-2021. The median portfolio size of 69 stocks and the median average allocation of 1.45\% indicate that most funds probably maintain well-diversified portfolios, aligning with the principles of modern portfolio theory.

Despite overall diversification, there are clear popularity trends among the stocks held by the funds. Table \ref{tab:explore_repeat_presence} presents the top 10 most popular stocks overall and the top 10 most popular stocks for exploration. The top overall stocks are dominated by well-established large-cap companies, while the top exploration stocks include a mix of companies from various sectors and countries (Table \ref{tab:explore_repeat_popular_desc}).

\begin{table}
\centering
\caption{Overall and Explore Percentage Presence of stocks in Portfolios for Q1-2021}
\begin{tabular}{llrlr}
\toprule
\multicolumn{1}{c}{} & \multicolumn{2}{c}{Top Overall} & \multicolumn{2}{c}{Top Explore} \\
\cmidrule(lr){2-3} \cmidrule(lr){4-5}
 & Ticker & \multicolumn{1}{c}{\%} & Ticker & \multicolumn{1}{c}{\%} \\
\midrule
1 & MSFT & 34.1 & TSLA & 5.6 \\
2 & AAPL & 28.2 & POOL & 4.2 \\
3 & AMZN & 27.3 & UNLYF & 3.7 \\
4 & GOOGL & 26.9 & GLT & 3.4 \\
5 & META & 26.7 & ETSY & 3.0 \\
6 & V & 26.0 & HLFFF & 2.8 \\
7 & UNH & 25.6 & ENPH & 2.5 \\
8 & JNJ & 23.8 & FNLPF & 2.2 \\
9 & CMCSA & 23.5 & MS & 2.2 \\
10 & MRK & 23.5 & KKOYF & 2.2 \\
\bottomrule
\end{tabular}
\label{tab:explore_repeat_presence}
\end{table}

\begin{table*}
\centering
\caption{Market Cap USD (2021-Q1), Sector and Country for top overall and top explore stocks}
\begin{tabular}{llll}
\hline
\multicolumn{4}{c}{Top Overall Stocks} \\
\hline
Ticker & Market Cap& Sector & Country \\
\hline
MSFT & 1876B & Technology & U.S. \\
AAPL & 2046B & Technology & U.S. \\
AMZN & 1806B & Consumer Cyclical & U.S. \\
GOOGL & 702B & Communication Services & U.S. \\
META & 722B & Communication Services & U.S. \\
V & 372B & Financial Services & U.S. \\
UNH & 369B & Healthcare & U.S. \\
JNJ & 395B & Healthcare & U.S. \\
CMCSA & 220B & Communication Services & U.S. \\
MRK & 178B & Healthcare & U.S. \\
\hline
\end{tabular}
\begin{tabular}{llll}
\hline
\multicolumn{4}{c}{Top Exploration Stocks} \\
\hline
Ticker & Market Cap& Sector & Country \\
\hline
TSLA & 719B & Consumer Cyclical & U.S. \\
POOL & 16B & Industrials & U.S. \\
UNLYF & 147B & Consumer Defensive & U.K. \\
GLT & 0.69B & Basic Materials & U.S. \\
ETSY & 24B & Consumer Cyclical & U.S. \\
HLFFF & 14B & Consumer Cyclical & Germany \\
ENPH & 19B & Technology & U.S. \\
FNLPF & 8B & Basic Materials & Mexico \\
MS & 135B & Financial Services & U.S. \\
KKOYF & 6B & Consumer Defensive & Finland \\
\hline
\end{tabular}
\label{tab:explore_repeat_popular_desc}
\end{table*}

Table \ref{tab:turnover} presents summary statistics for the turnover of portfolios across quarters. The mean turnover ranges from 5.7\% to 8.0\%, indicating that funds actively adjust their holdings over time. However, the wide range of values, from 0\% to 100\%, suggests that some funds maintain more stable portfolios while others undergo significant changes.

Portfolio turnover rates, summarized in Table \ref{tab:turnover}, highlight the ongoing adjustments made by funds to their holdings. The mean turnover ranges from 5.7\% to 8.0\%, with Q3-2020 exhibiting the highest mean turnover and standard deviation. The wide range of turnover rates suggests that some funds maintain more stable portfolios while others undergo significant changes.

\begin{table}
\centering
\caption{Summary statistics for portfolio turnover}
\begin{tabular}{lrrrr}
\toprule
 & Q2-2020 & Q3-2020 & Q4-2020 & Q1-2021 \\
\midrule
mean & 7.3\% & 8.0\% & 5.7\% & 6.2\% \\
std & 10.4\% & 11.4\% & 9.2\% & 10.2\% \\
min & 0.0\% & 0.0\% & 0.0\% & 0.0\% \\
25\% & 0.7\% & 1.2\% & 0.3\% & 0.5\% \\
50\% & 3.9\% & 4.5\% & 2.7\% & 3.0\% \\
75\% & 9.6\% & 9.7\% & 6.9\% & 7.6\% \\
max & 97.8\% & 100.0\% & 100.0\% & 100.0\% \\
\bottomrule
\end{tabular}
\label{tab:turnover}
\end{table}

The distribution of turnover rates is positively skewed, with the median consistently lower than the mean in all quarters. For instance, in Q1-2021, the median turnover is 3.0\%, while the mean is 6.2\%, suggesting that a few high-turnover funds are pulling the average up. The 25th and 75th percentile values show that turnover rates vary widely among funds, with some maintaining stable portfolios and others undergoing significant changes.

Notably, the maximum turnover rate of 100\% in Q3-2020, Q4-2020, and Q1-2021 indicates that some funds completely replaced their portfolio holdings during these quarters. These high-turnover funds present an opportunity for further research to understand their unique investment strategies or circumstances.

The exploratory data analysis reveals distinct patterns and characteristics that underscore the importance of analyzing repeat and explore portfolios separately. The insights gained from examining the dataset's structure, stock popularity trends, and portfolio turnover rates suggest that a nuanced approach to modeling and evaluation is necessary to effectively capture the dynamics of mutual fund investment decisions.

One of the key findings is the presence of clear popularity trends among the stocks held by funds. The top overall stocks are dominated by well-established large-cap companies, while the top exploration stocks exhibit more diversity in terms of market cap, sector, and country. This observation suggests that funds employ different strategies when maintaining their core holdings versus exploring new investment opportunities. Recommender systems that treat repeat and explore portfolios as homogeneous may fail to capture these distinct preferences and behaviors.

The insights gained suggest that a nuanced approach to modeling and evaluation is necessary to effectively capture the dynamics of mutual fund investment decisions and serve as a foundation for the next step in our research: apply recommender systems, including Next Basket Recommendation (NBR) techniques, to predict mutual fund portfolios, with a specific focus on the Next Novel Basket Recommendation (NNBR) task.

\section{Experimental Setting}

\subsection{Training and Validation Methodology}

\begin{figure}
    \includegraphics[width=\linewidth]{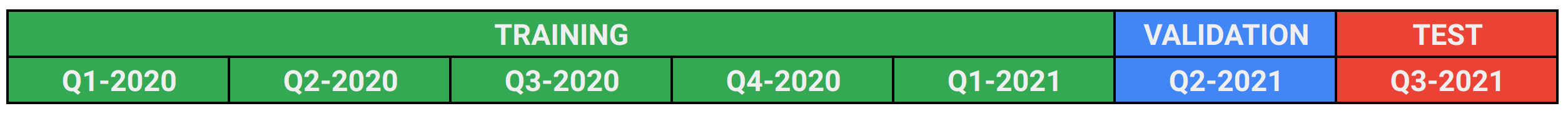}
    \caption{Train-Validation-Test temporal split}
    \Description[Train-Validation-Test temporal split]{Train-Validation-Test temporal split}
      \label{fig:train-val-test}
\end{figure}

To simulate real-world dynamics and predict the likely next purchases of funds in the upcoming quarter, we split the data into an out-of-time validation set, as represented by Figure~\ref{fig:train-val-test}. This time-based split is particularly important in financial applications, where the temporal aspect of investment decisions is crucial.

Sequential and NBR models use data from 2020 as historical data, with Q1-2021 being the target basket. For general recommenders, the entire 5-quarter period is summed and considered as a single interaction matrix. Further exploration of other options to model this temporal data is left for future research.

Unlike typical NBR papers that split data by users, we prioritize the temporal aspect of investment decisions over the ability to extrapolate to unseen funds. The more stable nature of mutual funds and stocks, when compared to other domains where NBR is usually applied, such as e-Commerce, allows us to focus on a time-based split while utilizing more general recommender system models that create user representations. The cold start problem in this context is less concerning, as our primary goal is to capture the temporal dynamics of mutual fund investment decisions rather than to predict the behavior of new funds.

To ensure reproducibility and test multiple modeling strategies, we used the RecBole recommendation library \cite{recbole}, a comprehensive, flexible, and efficient Python and PyTorch-based library that provides a unified framework for over 100 recommendation algorithms. The models tested include general recommendation models, sequential recommendation models, and two heuristic baselines (Global Asset Popularity and Current Allocation). We also evaluated Next Basket Recommendation models, which are not currently available in RecBole.

We utilized RecBole's benchmark file option to feed the models a predetermined train-validation-test split, ensuring a correct temporal split. For RecBole models, we used the atomic file data structure, focusing on .inter files representing historical user-item interactions.

General recommenders used an atomic file consisting of user\_id, item\_id, time\_field, and rating\_field. Data from the training period was merged, creating an entry for each fund-item interaction with the respective quarter and allocation. Models then converted this data to their respective representations, grouping interactions on the (user, item) level, irrespective of quarter, and converting the rating\_field to a binary value.

Sequential recommenders used a similar data structure, with the item\_id field as the target of a corresponding historical list of item interactions in the item\_id\_list field, leveraging the temporal nature of purchases.

For NBR models, we followed the data structures described by \cite{reality_check} and \cite{btbr}, organizing the data as a sequence of item sets fed to the models to predict the next set.

\subsection{Models}

We next describe the extensive set of techniques used in our experimental setting.

We evaluated a diverse set of techniques, including heuristic baselines, general recommenders, sequential models, and Next Basket Recommendation (NBR) models.

\subsubsection{Heuristic Baselines}
Two heuristic ranking methods, widely used as simple yet effective baselines in recommendation tasks [8], were employed to establish benchmarks for comparing the performance of the proposed models.

Global Asset Popularity ranks assets according to the number of unique funds that hold them in the quarter preceding the target quarter. It recommends the most widely held assets across all funds, similar to the global top-frequency metric (G-TopFreq) in the original paper.

Current Allocation assumes that funds are likely to maintain their current asset allocations. Analogous to the personal top-frequency (P-TopFreq) metric, it uses the fund's current allocation percentage for each asset to rank them, with higher allocations indicating a stronger preference. Assets already present in the portfolio are ranked above those outside.

\subsubsection{General Recommender System}
ItemKNN \cite{itemknn} is a nonparametric collaborative filtering model that computes item similarities based on the user-item interaction matrix. It generates recommendations by aggregating the similarities of the k most similar items for each user, without learning any parameters during training.

Neural Collaborative Filtering (NeuMF)\cite{neumf} combines matrix factorization (MF) and multilayer perceptron (MLP) models to capture complex user-item interactions. It uses separate embedding layers for users and items, which are then fed into MF and MLP components. 

EASE (Embarrassingly Shallow Autoencoders for Sparse Data) \cite{ease} is a linear model for collaborative filtering with sparse data. It calculates an item similarity matrix directly during initialization using a closed-form solution.

RecVAE (Recommender Variational Autoencoder) \cite{recvae} is an advanced collaborative filtering model that introduces a novel composite prior distribution for latent codes, a personalized approach to setting hyperparameters, and an alternating training strategy.

LightGCN is a simplified Graph Convolution Network (GCN) \cite{lgcn} model for collaborative filtering that removes the feature transformation and non-linear activation layers present in most GCN models. It focuses solely on neighborhood aggregation by linearly propagating user and item embeddings on the user-item interaction graph and using the weighted sum of embeddings learned at all layers as the final embedding.

GCMC (Graph Convolutional Matrix Completion) \cite{berg2017graph} represents user-item interaction data as a bipartite graph and employs a graph auto-encoder framework with differentiable message passing. The graph encoder produces node embeddings through graph convolutions, which are then fed into a bilinear decoder to predict unobserved ratings.

\subsubsection{Sequential Recommenders}

GRU4Rec\cite{gru4rec} is a sequential recommender model that incorporates Gated Recurrent Units (GRUs) to capture temporal dynamics and dependencies within user sessions. It takes item sequences as input, generates session-level representations using GRU layers, and predicts the next item in the session.

BERT4Rec\cite{bert4rec} is a sequential recommendation model based on the BERT (Bidirectional Encoder Representations from Transformers) architecture. It employs deep bidirectional self-attention to model user behavior sequences. BERT4Rec is trained using the Cloze objective, which involves predicting randomly masked items in the sequence by conditioning on their left and right context, allowing each item to fuse information from both sides using the attention mechanism.

\subsubsection{Next Basket Recommendation}
    
DNNTSP (Deep Neural Networks for Temporal Set Prediction) \cite{DNNTSP} is an integrated approach that combines graph neural networks and attention mechanisms to predict elements in a subsequent set given a sequence of sets. It learns element relationships by constructing set-level co-occurrence graphs and performing graph convolutions on these dynamic graphs. Additionally, DNNTSP employs an attention-based module to adaptively learn the temporal dependencies of elements and sets.

TIFUKNN (Temporal-Item-Frequency-based User-KNN) \cite{hu2020modeling} is a nearest neighbor-based method for next-basket recommendation that directly utilizes personalized item frequency (PIF) information. PIF records the number of times that each item is purchased by a user and provides critical signals for next-basket prediction. TIFUKNN models the temporal dynamics of frequency information in users' past baskets to introduce PIF, and then applies a k-nearest neighbors approach on the PIF data to generate recommendations. 

BTBR (Bi-directional Transformer Basket Recommendation) \cite{btbr} is a transformer-based model designed specifically for the Next Novel Basket Recommendation (NNBR) task, which focuses on recommending a basket consisting only of items that a user has not previously consumed. BTBR directly models item-to-item correlations within and across baskets, rather than learning complex basket representations. To effectively train BTBR, the authors propose several masking strategies and training objectives, such as item-level random masking, item-level select masking, basket-level all masking, basket-level explore masking, and joint masking.

It is worth noting that NBR models have certain limitations when dealing with large sets items, with undersampling being a common practice. In \cite{reality_check}, for example, they sample down users with baskets containing between 3 and 50 items. This is a particular problem in the mutual fund context, where the median portfolio size is 69 items. In our experiments, methods such as DNNTSP\cite{DNNTSP} would not run successfully when fed the entire portfolios, since they have quadratic runtime complexity when initializing their underlying data structures. This also required undersampling the funds baskets so that the algorithm would run in a manageable amount of time.

%\section{Evaluation Metrics and Tasks}

\subsection{Metrics and Tasks}

To evaluate the performance of the recommender models, we employ two widely used metrics in the recommender system domain: Recall@K and Normalized Discounted Cumulative Gain (NDCG@K), which assume that the recommended basket is of the same size K for all users.

Recall@K measures the ability to find all relevant items and is calculated as follows:
\begin{equation}
Recall@K(u_j) = \frac{|P_{u_j} \cap T_{u_j}|}{|T_{u_j}|},
\end{equation}
where $P_{u_j}$ is the predicted basket with K recommended items and $T_{u_j}$ is the ground truth basket for user $u_j$. The average recall score of all users is adopted as the recall performance.

NDCG@K is a ranking quality measurement metric that takes item order into account. It is calculated as follows:

\begin{equation}
NDCG@K(u_j) = \frac{\sum_{k=1}^{K} p_k/log_2(k+1)}{\sum_{k=1}^{min(K,|T_{u_j}|)} 1/log_2(k+1)},
\end{equation}
where $p_k$ equals 1 if $p^k_{u_j} \in T_{u_j}$, otherwise $p_k = 0$. $p^k_{u_j}$ denotes the k-th item in the predicted basket $P_{u_j}$. The average score across all users is the NDCG performance of the algorithm.

In contrast to Recall@K, which treats all items in the basket equally, NDCG@K assigns higher weights to items ranked higher in the predicted basket, making it more suitable for evaluating the relative ordering of items. This is particularly important in the context of basket recommendation, where the order of items within the basket may reflect user preferences or item importance.

Using these two metrics, we consider three distinct tasks for the next portfolio prediction, inspired by the research of \cite{btbr} on next basket recommendation (NBR). These tasks are defined as follows:

\begin{enumerate}
\item \textbf{Next Basket Recommendation (NBR):} The general task of predicting the next basket of items, without distinguishing between repetition and exploration items. This task aims to recommend all possible items that a user might purchase in their next basket, based on their historical basket sequence. In the context of portfolio prediction, this task aims to predict the entire composition of the next portfolio for each mutual fund, considering both new and previously held stocks.

\item \textbf{Next Novel Basket Recommendation (NNBR):} This task focuses on recommending a novel basket, i.e., a set of items that are new to the user, given the user's historical basket sequence. NNBR is particularly relevant to the concept of exploring new investment opportunities in the portfolio management scenario. It could be used to help mutual funds discover new potential stocks that align with their investment strategies and historical portfolio compositions.

\item \textbf{Next Basket Repurchase Recommendation (NBRR):} This task focuses on the pure repetition task, i.e., recommending items that users have purchased before, based on their historical basket sequence. NBRR is relevant to the concept of maintaining existing investments in the portfolio management context. It aims to predict the stocks that a mutual fund is likely to continue holding in its next portfolio, based on its historical investment patterns.
\end{enumerate}

The two metrics described, Recall@K and NDCG@K, are applied to evaluate the performance of the recommender models in each of these three tasks. However, for the NNBR and NBRR tasks, the item universe is filtered a priori, meaning that only novel items are considered for NNBR and only repeat items are considered for NBRR.  For the NNBR task, let $N_{u_j}$ denote the set of novel items for user $u_j$, i.e. items with which the user has not interacted in the past (in our case, we consider the past 4 quarters). The Recall@K and NDCG@K formulas for the NNBR task are simply modified by only considering the $N_{u_j}$ set of items. Similarly, for the NBRR task, let $R_{u_j}$ denote the set of repeat items for user $u_j$, that is, items that the user has interacted with in the past (again, we consider the last 4 quarters. The Recall@K and NDCG@K formulas are modified for the NBRR task by only considering the $N_{u_j}$ universe of iteractions. 

In both NNBR and NBRR tasks, the average scores across all users are used as the final performance metrics for the recommender models. By applying these modified formulas, the evaluation focuses solely on the relevant items for each task (novel items for NNBR and repeat items for NBRR), ensuring a fair and accurate assessment of the models' performance in these specific scenarios.

To further ensure reproducibility, we use the torchmetrics\cite{torchmetrics} implementation of both Retrieval Recall and Retrieval Normalized Discounted Cumulative Gain. Extending these implementations to support NNBR and NBRR is straightforward: we simply multiply both targets and prediction by a binary exploration matrix in the case of NNBR or its boolean inverse in the case of NBRR.

By considering these distinct tasks and evaluating them using appropriate metrics, we aim to provide a comprehensive analysis of the models' performance in various scenarios related to next portfolio prediction, taking into account both item retrieval and ranking quality. This approach allows us to assess the models' ability to predict the entire portfolio composition (NBR), identify new investment opportunities (NNBR), and maintain existing investments (NBRR) for mutual funds, based on their historical portfolio data.

\section{Results and Discussion}

\begin{table*}[t]
\caption{NBR (all items), NBRR (repurchase items) and NNBR (novel items) evaluation metrics}
\medskip
\small
Validation and test results for models trained with data from Q1-2020 through Q1-2021\\
Best model in \textbf{bold} and runner-up \underline{underlined}, statistically significant with a 95\% CI
\begin{tabular}{lrrrrrrrrrrrr}
\toprule
 & \multicolumn{2}{c}{NBR recall@20} & \multicolumn{2}{c}{NBR ndcg@20} & \multicolumn{2}{c}{NBRR recall@20} & \multicolumn{2}{c}{NBRR ndcg@20} & \multicolumn{2}{c}{NNBR recall@20} & \multicolumn{2}{c}{NNBR ndcg@20} \\
 & Q2-2021 & Q3-2021 & Q2-2021 & Q3-2021 & Q2-2021 & Q3-2021 & Q2-2021 & Q3-2021 & Q2-2021 & Q3-2021 & Q2-2021 & Q3-2021 \\
\midrule
DNNTSP & \underline{0.337} & \underline{0.320} & \underline{0.939} & \underline{0.893} & \underline{0.367} & \underline{0.342} & \underline{0.944} & \underline{0.897} & 0.028 & 0.032 & 0.027 & 0.038 \\
TIFUKNN & 0.316 & 0.304 & 0.899 & 0.866 & 0.347 & 0.328 & 0.906 & 0.873 & 0.099 & 0.101 & 0.084 & 0.085 \\
BTBR & 0.013 & 0.017 & 0.093 & 0.111 & 0.002 & 0.051 & 0.065 & 0.327 & 0.157 & 0.090 & 0.115 & \underline{0.111} \\
\hline
EASE & 0.258 & 0.251 & 0.806 & 0.783 & 0.332 & 0.324 & 0.871 & 0.860 & \textbf{0.210} & \textbf{0.159} & \textbf{0.157} & \textbf{0.118} \\
RecVAE & 0.261 & 0.253 & 0.777 & 0.751 & 0.329 & 0.323 & 0.856 & 0.847 & \underline{0.193} & \underline{0.154} & \underline{0.145} & 0.110 \\
LightGCN & 0.214 & 0.208 & 0.677 & 0.654 & 0.327 & 0.322 & 0.843 & 0.833 & 0.179 & 0.137 & 0.133 & 0.102 \\
GCMC & 0.193 & 0.189 & 0.641 & 0.623 & 0.329 & 0.323 & 0.850 & 0.842 & 0.169 & 0.138 & 0.130 & 0.100 \\
NeuMF & 0.211 & 0.205 & 0.623 & 0.601 & 0.320 & 0.315 & 0.810 & 0.808 & 0.169 & 0.120 & 0.127 & 0.088 \\
GRU4Rec & 0.212 & 0.207 & 0.699 & 0.681 & 0.343 & 0.332 & 0.895 & 0.875 & 0.162 & 0.127 & 0.119 & 0.095 \\
ItemKNN & 0.156 & 0.154 & 0.603 & 0.588 & 0.316 & 0.311 & 0.849 & 0.840 & 0.140 & 0.124 & 0.114 & 0.099 \\
BERT4Rec & 0.012 & 0.012 & 0.095 & 0.095 & 0.250 & 0.250 & 0.735 & 0.740 & 0.010 & 0.022 & 0.008 & 0.014 \\
\hline
Last Alloc. & \textbf{0.342} & \textbf{0.348} & \textbf{0.956} & \textbf{0.958} & \textbf{0.374} & \textbf{0.372} & \textbf{0.961} & \textbf{0.962} & 0.002 & 0.003 & 0.004 & 0.004 \\
Popularity & 0.053 & 0.051 & 0.256 & 0.250 & 0.325 & 0.319 & 0.811 & 0.810 & 0.050 & 0.039 & 0.038 & 0.033 \\
Expl. Pop. & 0.015 & 0.018 & 0.117 & 0.119 & 0.314 & 0.316 & 0.783 & 0.791 & 0.037 & 0.035 & 0.032 & 0.033 \\
Random & 0.004 & 0.004 & 0.037 & 0.037 & 0.308 & 0.309 & 0.772 & 0.784 & 0.005 & 0.005 & 0.004 & 0.005 \\
\bottomrule
\end{tabular}
\label{tab:metrics}
\end{table*}

Table \ref{tab:metrics} summarizes the performance of various models in the NBR, NBRR, and NNBR tasks. The results reveal intriguing patterns and insights into the effectiveness of different approaches for mutual fund portfolio prediction.

For both NBR and NBRR tasks, which consider the entire portfolio and repeat-only stocks respectively, the current allocation heuristic consistently outperforms all other models in terms of both recall and NDCG. 

This simple yet effective approach highlights the inherent stability in mutual fund portfolios, suggesting that fund managers tend to maintain a significant portion of their holdings over time.

NBR-specific models DNNTSP and TIFUKNN secure the second and third positions, respectively. While these models are indeed optimized for their task, their inability to surpass the current allocation heuristic underscores the challenge of capturing the nuanced decision-making processes of fund managers through more complex algorithms.

The NNBR task, focusing on predicting novel items in portfolios, presents a markedly different landscape. Here, autoencoder-based models, particularly EASE and RecVAE, demonstrate superior performance. The success of EASE, a nonparametric model, in outperforming more complex techniques is particularly noteworthy. This suggests that the latent patterns in mutual fund exploration behavior might be effectively captured through simpler and interpretable models.

LightGCN's strong performance indicates that graph-based approaches can effectively model the complex relationships between funds and stocks in the context of novel item prediction. BTBR, despite being custom built for the NNBR task, shows improvement over other NBR models, but falls short of the top-performing techniques. This highlights the challenge of developing specialized models that can outperform more general approaches in this domain.

Comparing model categories reveals interesting trends. General recommenders excel in the NNBR task, while sequential models perform moderately well on tasks without being compared to each other. NBR-specific models show strength in NBR and NBRR tasks but struggle with novel item prediction. Graph-based models demonstrate balanced performance, indicating their potential in capturing complex fund-stock relationships.

It is important to note that the NBR and NBRR NDCG metrics for the current allocation heuristic are 0.956 and 0.961, respectively, while the NDCG for the best NNBR model, EASE, is only 0.21. Since NDCG is a normalized metric between 0 and 1, this further highlights the inherent difficulty of the NNBR task and indicates that there is significant room for improvement in future models.
Additionally, it is worth mentioning that we used the default implementation and hyperparameters for all models tested. Further refinements and optimizations in this area could potentially lead to improved performance across all tasks.

These findings have several practical implications for the mutual fund industry:

The strong performance of the current allocation heuristic in NBR and NBRR tasks suggests that simple, interpretable models can be highly effective for predicting the core holdings of mutual funds. This could be valuable for investors seeking to replicate fund strategies or for regulatory bodies monitoring fund behavior.

The success of auto-encoder-based models in the NNBR task indicates their potential to identify emerging trends or shifts in investment strategies. Fund managers could leverage these insights to stay ahead of market trends or identify potential investment opportunities that align with their fund's objectives.

The balanced performance of graph-based models across tasks suggests their potential as versatile tools for comprehensive portfolio analysis and prediction.

A notable observation is the considerable degradation of performance in the test set for all models, although their relative ranking remains mostly stable, particularly among top performing ones. This degradation highlights the dynamic nature of mutual fund strategies and market conditions.
Several factors could contribute to this performance drop, such as market volatility and changing economic conditions between training and test periods, changes in fund strategies or management changes, and the introduction of new financial instruments or changes in regulatory environments, among others.

This observation underscores the need for frequent model retraining in real-world applications to capture the latest market trends and fund behaviors. It also highlights the challenge of developing models that can maintain performance in the face of evolving financial landscapes.
% \subsection{Limitations and Challenges}
% Despite the insights gained, several limitations and challenges are worth noting:

% Data limitations: The dataset covers a relatively short time period, which may not capture long-term trends or cyclical behaviors in mutual fund strategies.

% Model complexity: While more complex models like DNNTSP show strong performance in some tasks, they are easily outperformed by simpler models such as EASE or the current allocation heuristic in others. This suggests that increasing model complexity does not necessarily lead to better performance in this domain.

% Handling of market conditions: The current models don't explicitly account for broader market conditions or economic indicators, which could significantly influence fund decisions.

\section{Conclusion}

In this work, we considered the task of predicting the next quarter portfolio allocation of a mutual fund given historic data on its previous allocation, a task that has been overlooked in the literature.
We formulated the task as an instance of Next Novel Basket Recommendation (NNBR), assembled a benchmark using publicly available data and carried out experiments comparing several state-of-the-art techniques.

Our findings showed that, overall, even sophisticated NBR models are outperformed by very simple heuristics such as repeating the last allocation.
When considering only the more challenging task of predicting newly acquired items, autoencoder-based recommendation techniques that ignore the temporal and basket information outperformed all other techniques.

This study has significant practical implications for both fund managers and investors in the mutual fund industry. By leveraging our insights, fund managers can enhance their investment decision-making processes, identify new opportunities, and stay ahead of market trends. In particular, by using autoencoder-based models, fund managers can adapt to changing market conditions, optimize portfolio allocations, and make more informed investment decisions.
%, ultimately leading to improved fund performance and better outcomes for investors.
From an investor's perspective, this study contributes to increase transparency in the mutual fund industry and enable more informed decision-making when selecting funds. The development of tools that leverage the best-performing models for portfolio prediction can empower investors to compare and evaluate mutual funds based on their predicted allocations, aligning their investment choices with their financial goals and risk preferences. 

The benchmark dataset created in this study will potentially serve as a valuable resource for future research and development efforts in the field of AI for finance, fostering collaboration and knowledge sharing among researchers and practitioners to accelerate the development of sophisticated and tailored models for mutual fund portfolio prediction and other financial applications.

Finally, from a scientific viewpoint, this study underscores the importance of considering domain-specific characteristics when applying recommendation  techniques to financial applications.

%In conclusion, this study makes a significant contribution to the field of AI applications in finance by introducing a novel application of NNBR techniques to mutual fund portfolio prediction and providing a comprehensive evaluation of various recommender system models. The insights gained have practical implications for fund managers and investors, offering the potential to enhance investment decision-making, increase transparency, and ultimately lead to better outcomes for all stakeholders in the mutual fund industry. As the adoption of AI and machine learning techniques continues to grow in the financial sector, the findings of this study will serve as a foundation for future research and development efforts, driving innovation and shaping the future of AI-driven investment management.

\bibliography{acmart}

\end{document}